# Tunable Quantum Spin Hall Effect via Strain in two-Dimensional Arsenene Monolayer

Ya-ping Wang, Chang-wen Zhang*, Wei-xiao Ji, Run-wu Zhang, Ping Li, Pei-ji Wang, Miao-juan Ren, Xin-lian Chen, and Min Yuan

School of Physics and Technology, University of Jinan, Jinan, Shandong, 250022, People's Republic of China

**Abstract:** The search for new quantum spin Hall (QSH) phase and effective manipulations of their edge states are very important for both fundamental sciences and practical applications. Here, we use first-principles calculations to study the strain-driven topological phase transition of two-dimensional (2D) arsenene monolayer. We find that the band gap of arsenene decreases with increasing strain and changes from indirect to direct, and then the s-p band inversion takes place at Γ point as the tensile strain is larger than 11.14%, which lead to a nontrivially topological state. A single pair of topologically protected helical edge states is established for the edge of arsenene, and their QSH states are confirmed with nontrivial topological invariant $Z_2 = 1$. We also propose high-dielectric BN as an ideal substrate for the experimental synthesis of arsenene, maintaining its nontrivial topology. These findings provide a promising candidate platform for topological phenomena and new quantum devices operating at nanoelectronics.

* Corresponding author: C. W. Zhang: zhchwsd@163.com TEL: 86-531-82765976

Two-dimensional (2D) topological insulators (TIs), also known as the quantum spin Hall (QSH) insulators,[1–7] are characterized by an insulating bulk and gapless edge states at its boundaries.[1, 2] These edge states, which are spin-locked due to the protection of time-reversal symmetry (TRS), are topologically protected from backscattering of non-magnetic defects or impurities, thus leading to dissipationless transport edge channels for novel quantum electronic devices.[6, 7] The prototypical concept of QSH insulator is first proposed by Kane and Mele in graphene,[1, 2] in which the spin-orbit coupling (SOC) opened a band gap at the Dirac point. However, the associated gap is too small (~$10^{-3}$ meV), which makes the QSH effect in graphene only appear at an unrealistically low temperature.[8-10] Quantized conductance through QSH edge states have been experimentally demonstrated in HgTe/CdTe[4, 5] and InAs/GaSb[11, 12] quantum wells. Such systems have stimulated enormous research activity due to their novel QSH effect and hence the potential application in quantum computation and spintronics[13, 14].

Group-V elemental monolayers have attracted interests as novel 2D materials with semiconducting electronic properties. For example, the monolayer form of black phosphorous, phosphorene (α-P), has been reported to have a direct band gap and high carrier mobility, which can be exploited in the electronics[15,16]. Besides, Bi or Sb ultrathin films[17-20], the 2D group-V honeycomb-like materials with the strongest SOC effect, have been proposed to host large-gap QSH insulators, ample to applications at room temperature. Recently, arsenene in α and β phases has also been reported to be energetically stable[21-24]. These materials with high mechanical stretchability, which can reversibly withstand extreme mechanical deformation and cover arbitrary surfaces and movable parts, are used for stretchable display devices, broadband photonic tuning and aberration-free optical imaging. However, the topological properties of honeycomb-like arsenene have not been studied up to date. It is thus reasonable to ask whether or not the arsenene can become a nontrivial QSH insulator, which maybe largely enhances its application in spintronic devices.

In this work, based on first-principles calculations, we predict a new QSH insulator in a 2D buckled arsenene via in-plane strain. Noticeably, at a critical value of tensile strain of 11.14%, band inversion occurs at Γ point, causing a topological phase transition from a trivial to a non-trivial QSH insulator. A single pair of topologically protected helical edge states is established for arsenene, and its QSH states are confirmed with nontrivial topological invariant $Z_2 = 1$. We also propose high-dielectric BN as an ideal substrate for the experimental realization of arsenene, maintaining its nontrivial topology. These findings are promising platforms for topological phenomena and possible applications in spintronics.

First-principles calculations based on density-functional theory (DFT)[25] are performed by the Vienna ab initio simulation package[26], using the projector-augmented-wave potential. The exchange-correlation functional is treated using the Perdew-Burke-Ernzerhof (PBE)[27] generalized-gradient approximation. The energy cutoff of the plane waves is set to 600 eV with the energy precision of $10^{-5}$ eV. The Brillouin zone (BZ) is sampled by using a 9×9×1 Gamma- centered Monkhorst–Pack grid, and the vacuum space is set to 20 Å to minimize artificial interactions between neighboring slabs. All structures are fully optimized, including cell parameters and atomic coordinates, until the residual forces are less than 0.001 eV/Å. The SOC is included in the self-consistent calculations of electronic structure.

Bulk As has four allotropes, and the most stable one is gray As[24], which is rhombohedral with two atoms per primitive cell. Thus, it can be viewed as a stacking of the bilayers along the [111] direction, as shown in Fig. 1(a). Unlike the plannar graphene, the peeled arsenene monolayer has a buckled honeycomb structure, with an optimized buckling distance h = 1.39 Å, bond length d = 2.51 Å, as well as an angle of θ = $33.7^0$ (Fig. 1(b)), in consistent with that of Ref. 24. Generally, the buckle configuration can sustain a larger mechanical strain than planner one, thus its structural evolution can be realized by the external in-plane strain. To well understand why arsenene can suffer from such a large strain without dissociation, we analyze the variation of buckled height (h), bond-length (d) and the bond-angle (θ) via an external axial strain. Figs. 1 (c) and (d) display the changes of in-plane bond-length and inter-plane bond-angle in arsenene monolayer. One can see that, under the tensile strain, both the bond-length and buckled height change slightly with respect to the bond-angle variations. For example, when imposing 12% tensile strain, the value of d (h) changes by 6.13 % (8.47 %), but θ are reduced significantly by 15.16%, much larger than the relative variations of d or h. Thus, the change of bond-angle greatly relieves the variations of bond-length and makes arsenene being stable even under large magnitude of mechanical strain. The dynamic stability is also checked with the phonon spectrum calculated along the highly symmetric directions for all monolayers. No modes with imaginary frequencies in the spectrum are found, thus these monolayers are expected to be dynamically stable. These structural evolutions play a crucial role in the engineering of band structures in arsenene monolayer.

Figure 2 displays the band structures of arsenene with respect to external tensile strain. One can see that, at the equilibrium state, it is indirect-gap semiconductor (Fig. 2 (a)), with its valence band maximum (VBM) located at Γ point and conduction band minimum (CBM) on M-Γ path. When increasing the in-plane strain, its band structure gradually become direct-gap type, with both the CBM and VBM at Γ point, as displayed in Fig. 2 (b). Further increasing the strain, we find that the CBM is driven continuously to shift downward to the Fermi level, while the VBM increases reversibly, leading the band gap to decrease monotonically (Fig. 2 (c)). Notably, at the critical value of 11.14 %, the two bands touch each other at the Fermi level, which can be considered as a semi-metal with zero density of states (Fig. 2 (d)). In Fig. 3(b), we displays 3D band structure around the Fermi level imposing on 11.14 % tensile strain, further indicating band feature near the Fermi level with upper conduction band linearly separated. However, it is interesting to find that the band gap reopens again if the strain is larger than 11.14%. In this case, the CBM is transformed from the "−shape" to "M-shape", while the VBM remains flattened at Γ point, as illustrated in Figs. 2 (e) and 2(f). In another word, arsenene becomes an indirect-gap semiconductor again. The band gap at the Γ point as a function of strain is plotted in Fig. 3(a), which shows that the system undergoes a gap closing and reopening process, indicating the possibility of forming a QSH phase.

To further identify the nontrivial band topology in 2D arsenene, we calculate the $Z_2$ invariants ν following the approach proposed by Fu and Kane,[28] due to the presence of structural inversion symmetry. Here, the invariants $v$ can be derived from the parities of wave function at the four time-reversal-invariant momenta (TRIM) points $K_i$, namely one Γ point and three equivalent M points in the Brillouin zone, as illustrated in the insert of Fig. 3(a). Accordingly, the topological indexes $v$ are established by

$$\delta(K_i) = \prod_{m=1}^{N} \xi_{2m}^{i}, (-1)^{v} = \prod_{i=1}^{4} \delta(K_i) = \delta(\Gamma)\delta(M)^3 \tag{1}$$

where $\delta$ is the product of parity eigenvalues at the TRIM points, $\xi = \pm 1$ are the parity eigenvalues and $N$ is the number of the occupied bands. According to the $Z_2$ classification, $v = 1$ characterizes a QSH insulator, whereas $v = 0$ represents a trivial band topology. As expected, in the equilibrium state, the products of the parity eigenvalues at these two symmetry points: Γ(0.0, 0.0) and M(0.5, 0.5) are both -1, while at the M(0.0, 0.5) and M(0.5, 0.0) displays +1, yielding a trivial topological invariant $Z_2 = 0$. However, if the strain increases beyond 11.14 %, s-p band order inversion at Γ point takes place. It can be seen in Fig. 2, the s-type orbital confined to the CBM gradually shifts down toward VBM at Γ point with increasing strain, and then the valance and conduction bands touch each other and the s-type character is acquired by VBM. Such s-p band order exchange lead the parity eigenvalue of the VBM changes sign from - to +, while those at the M(0.5, 0.0), (0.0, 0.5), (0.5, 0.5) points remain +, +, -, respectively. Accordingly, the products of the parity eigenvalues at these points are now distinct and the system becomes TI with $Z_2 = 1$.

The SOC-induced gap opening near the Fermi level indicates possible existence of 2D TI state that are helical with the spin-momentum locked by TRS. To check this, we calculated the topological edge states of arsenene by the Wannier90 package[29] .Using DFT bands as input, we construct the maximally localized Wannier functions (MLWFs) and fit a tight-binding Hamiltonian with these functions. Fig. 3(c) shows the DFT and MLWFs fitted band structures of 12% tensile arsenene, which are in very good agreement with each other. Then, the edge Green's function[30] of a semi-infinite arsenene is constructed and the local density of state (LDOS) of as zigzag edge is calculated, as shown in Fig. 3(d). Clearly, all the edge bands are seen to connect the conduction and valence bands and span the 2D bulk energy gap, yielding a 1D gapless edge states. Besides, the counter-propagating edge states exhibit opposite spin polarizations, in accordance with the spin-momentum locking of 1D helical electrons. Furthermore, the Dirac point located at the band gap are calculated to have a high velocity of ~ $1.0 \times 10^5$ m/s, comparable to that of $5.5 \times 10^5$ m/s in HgTe/CdTe quantum well.[4,5] All the above results consistently indicate that arsenene is an ideal 2D TI.

The substrate materials are inevitable in device application, thus a free-standing film must eventually be deposited or grown on a substrate. Previous works indicate that the nontrivial TI features of graphene, silicene, and germanene[31-36] are easily destroyed by the substrate, thus introducing a trivial gap. In contrast, although the TI feature of arsenene monolayer are for free-standing structure, their nontrivial QSH would be quite robust when they are on the substrate, because their band inversion occurs at Γ point rather than K point, as well as the full saturation of As-$p_z$ orbitals ensures a weak interaction with the substrate. To check this idea, we select BN as a substrate with 12% tensile arsenene to form As/BN heterostructure, as shown in Fig. 4(a-b). For the structural relaxation, van der Waals (vdW) forces[37] are included in the calculations. The calculated binding energy is found to be -78 meV, indicating that it is a typical vdW heterostructure. The optimized lattice constant of arsenene is 4.20 Å, fall in the range of TI feature, maintaining the buckled height (h) and interlayer distance (d) being 1.12 Å and 3.27 Å, respectively. In this case, it has a small lattice mismatch (~3.68 %) in comparison to BN ($\sqrt{3} \times \sqrt{3}$) substrate, showing that it is feasible to grown arsenene on BN substrate. As expected from the band structure with SOC in Fig. 4(c), the As/BN heterostructure remains semiconducting. There is essentially no charge transfer between adjacent layers, thus the states around $E_F$ are dominantly contributed by arsenene. In comparison to the free-standing arsenene, little difference is observed between them. Evidently, As/BN heterostructure is also a nontrivial TI whose s-p band inversion is preserved, suggesting its robust QSH effect.

In summary, we demonstrate a strain-induced topological transition in 2D arsenene monolayer, accompanying by a band inversion that causes the change in the topological invariant from $Z_2 = 0$ to $Z_2 = 1$. Interestingly, the topological phase transition in arsenene is closely related to the changes of bonding angles, which would be tuned drastically due to the bonding strength changes. The QSH features of 2D topological insulators characterized by an explicit demonstration of the topological helical Dirac type edge states. In addition, the high-dielectric BN can be as an ideal substrate for the experimental synthesis of arsenene, maintaining its nontrivial topology. These results are expected to stimulate further work to synthesize, characterize and utilize arsenene for fundamental exploration and practical applications in spintronics.

___________________________

**Acknowledgments:** We would like to thank Hongbin Zhang for useful discussions. This work was supported by the National Natural Science Foundation of China (Grant No. 11274143, 61172028, and 11304121), and Research Fund for the Doctoral Program of University of Jinan (Grant No. XBS1433).

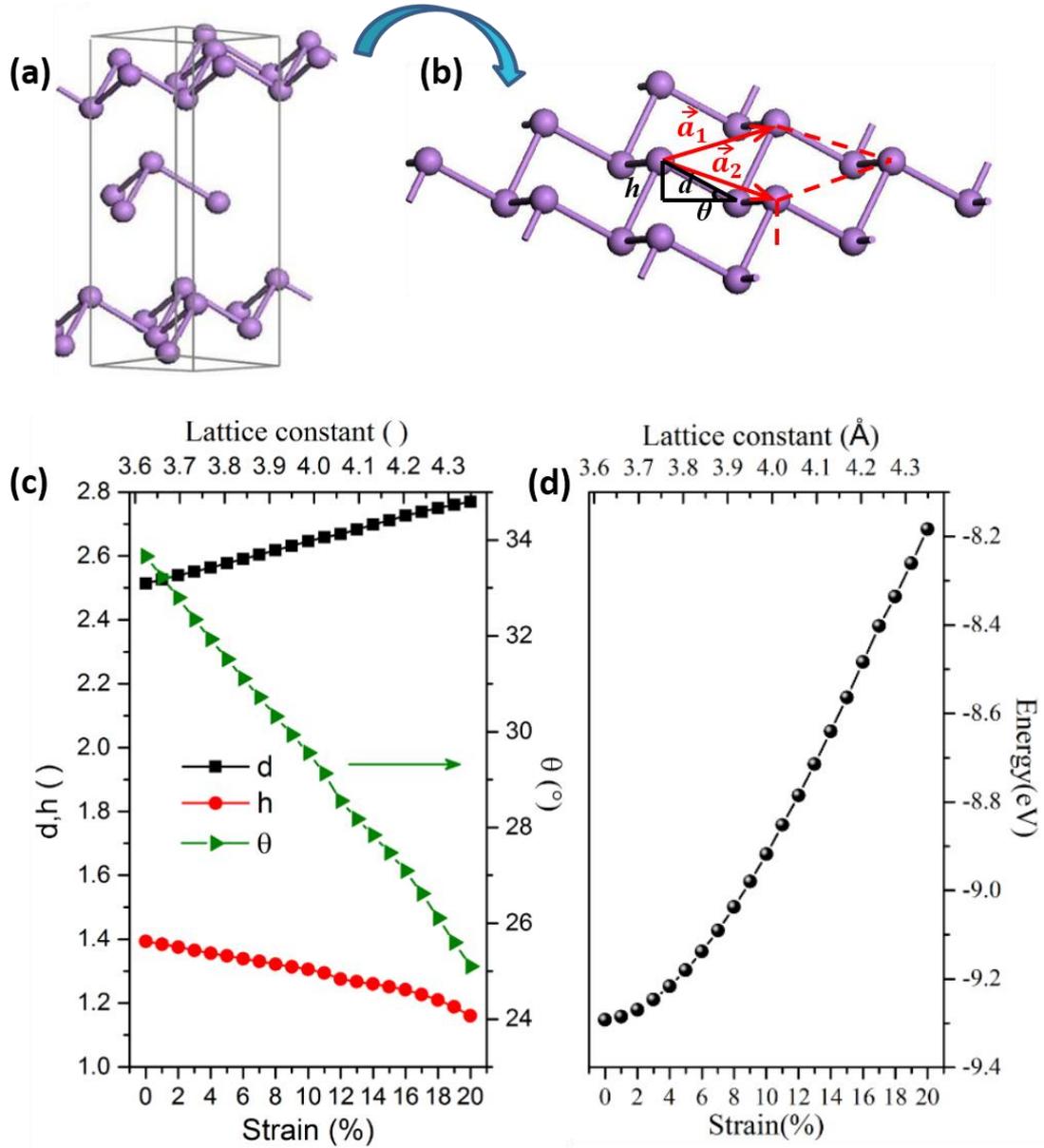

**Fig. 1** (a) Side view of the rhombohedrally (ABC) stacked layered structure of bulk gray arsenic. (b) The buckled honeycomb structure of a gray arsenic monolayer. (c) The variations of the buckled height (h), bond-length (d) and the bond-angle (θ) with the application of an axial strain. (d) The function of energy with axial stress.

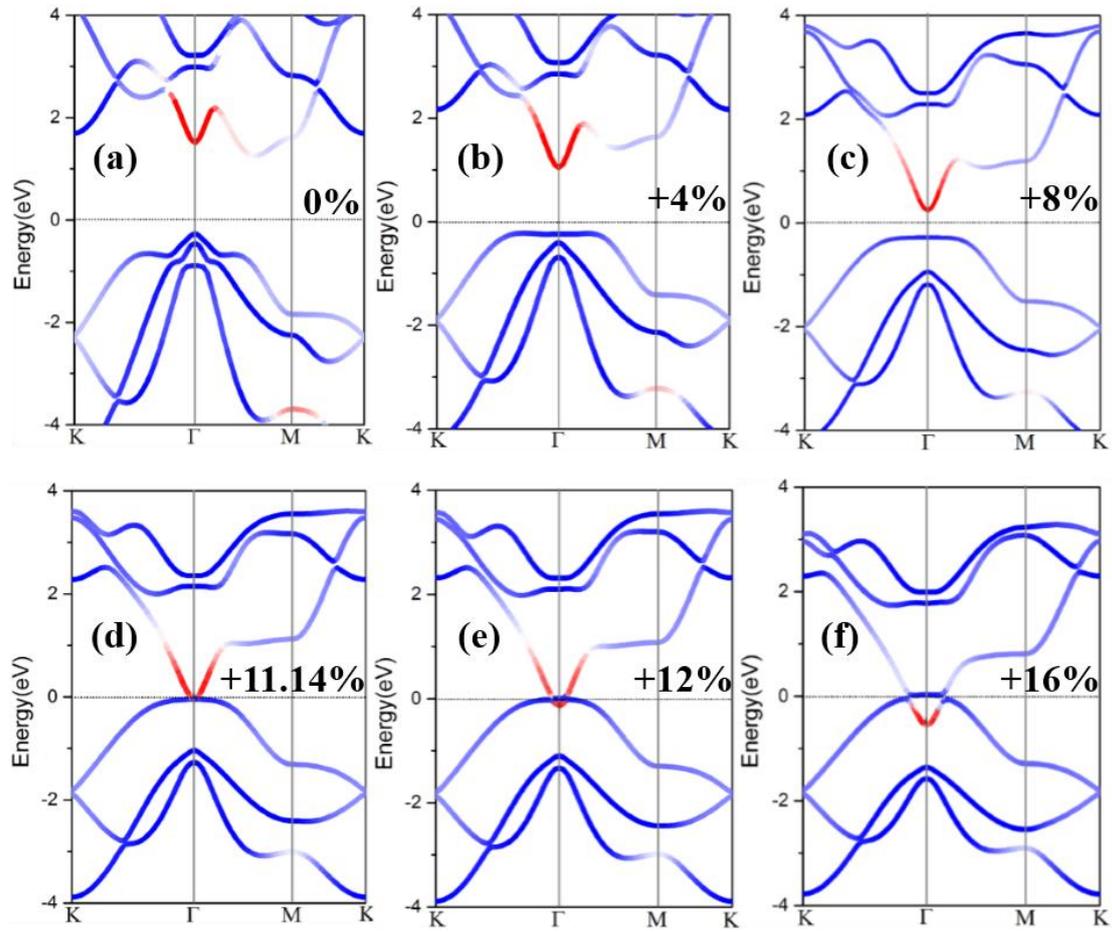

**Fig.2** (a-f) The band structures of arsenene with respect to external tensile strain, while the red section denotes the contribution of s-type orbital and blue section means the contribution of p-type orbital.

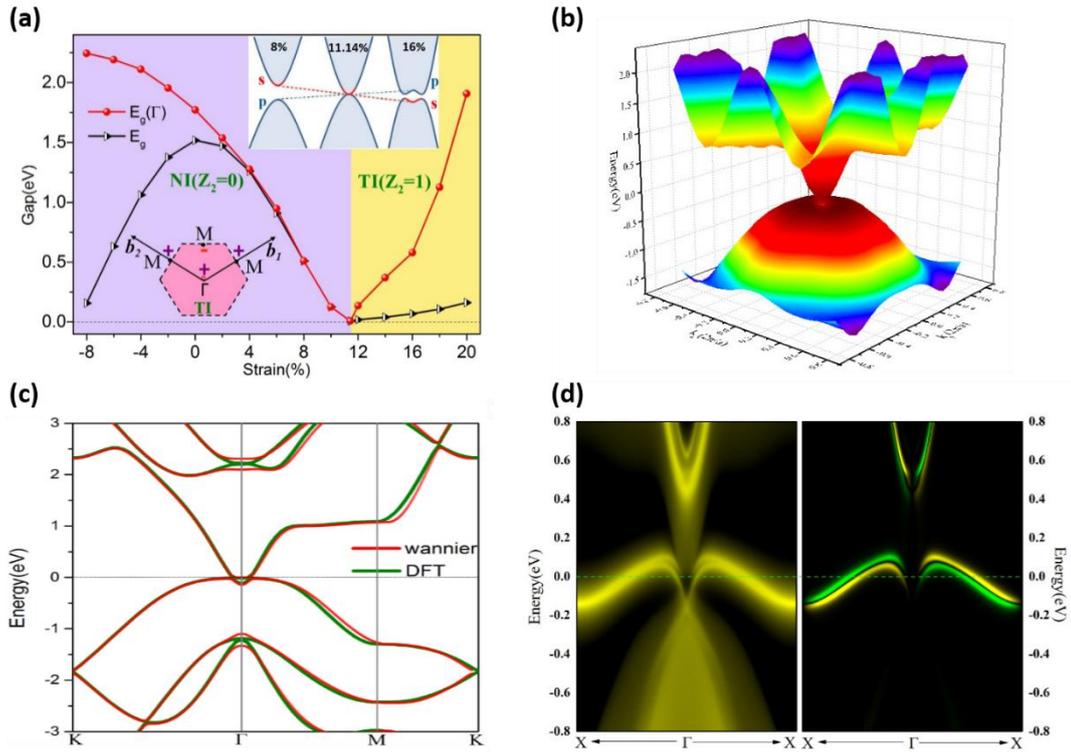

**Fig.3** (a) The variations of band gap with respect to external strain, along with the schematic of s-p band reverse in Brillouin zone in the insert, (b) The 3D band structure of 11.14% strain, (c) DFT and MLWFs fitted band structures of 12% tensile arsenene. (d) Electronic structure of helical edge states of 12% tensile strain, the left subpanel shows the total density of states while the right subpanel shows the corresponding spin polarization in two channels.

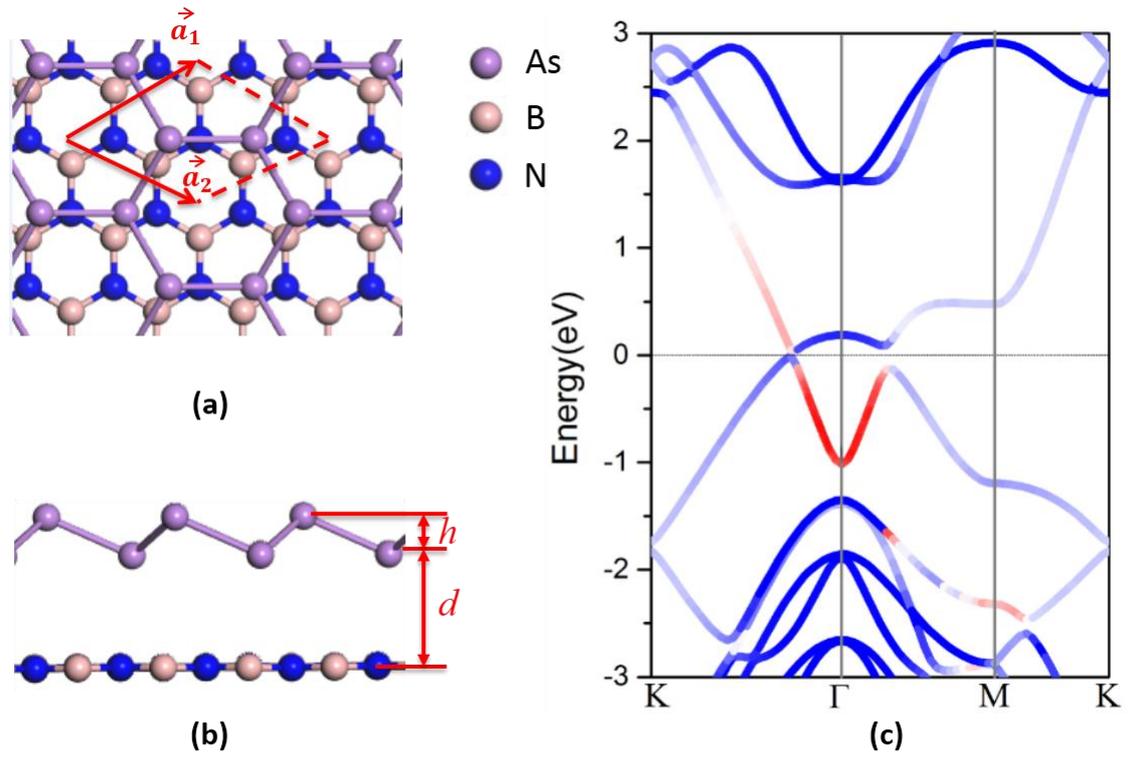

**Fig. 4** (a) Top and (b) side view of arsenene/BN heterostructure. (c) The band structures of the heterostructure.